# Origin of Life

Ashwini Kumar Lal

Ministry of Statistics & Programme Implementation
New Delhi, India

ashwinikumar_lal@yahoo.co.in

## Abstract

The evolution of life has been a big enigma despite rapid advancements in the fields of biochemistry, astrobiology, and astrophysics in recent years. The answer to this puzzle has been as mind-boggling as the riddle relating to evolution of Universe itself. Despite the fact that panspermia has gained considerable support as a viable explanation for origin of life on the Earth and elsewhere in the Universe, the issue remains far from a tangible solution. This paper examines the various prevailing hypotheses regarding origin of life like abiogenesis, RNA world, iron-sulphur world, and panspermia; and concludes that delivery of life-bearing organic molecules by the comets in the early epoch of the Earth alone possibly was not responsible for kick-starting the process of evolution of life on our planet.

## 1.    Introduction :

The question of evolution of life on the Earth and elsewhere in the Universe has ever been as challenging as the question of evolution of the Universe itself. Science does not provide authentic explanation regarding the origin of Universe in the controversial 'Big Bang' theory for evolution of the Universe (Arp et al. 1990), nor does it provide any satisfactory explanation regarding the origin of life despite considerable advancements in the fields of astrobiology, astrophysics, and molecular biology in recent years. The hot inflationary 'Big Bang' model for evolution of the Universe is not secure enough to serve as a foundation for beliefs about the origin of life, which is exemplified very much by the fact that the most distant galaxies, we can see today, look as rich and fully evolved as our own, even though they are barely 5% of the current age of the Universe, i.e. about 700 million years after the Big Bang (z~7), as revealed in the Hubble Ultra Deep Field (HUDF) images taken with Hubble's Advance Camera for Surveys and Near Infrared Camera. Among the several factors leading to beginning of life on this planet, 'panspermia' appears to provide the most favoured hypothesis for emergence of life on our planet. This paper examines the various prevailing hypotheses regarding origin of life on this planet. It also hints at a very interesting and crucial inference that probably delivery of life-bearing organic molecules by the comets in the early history of Earth alone was not sufficient to provide the requisite trigger mechanism for initiation of life on our planet.

## 2. Early Earth and Beginning of Life :

Our Sun and solar system were created from the nebular debris spawned by a red giant star which exploded in a supernova, nearly 5 billion years ago. The entire solar system including Earth is believed to have been created within 100 million years of this explosion. The current age of the Earth is estimated to be 4.6 Gyr. The nascent Earth was quite different from the world known today. There were no oceans and oxygen in the atmosphere. During the period 4.6 - 3.8 Gya (the Hadean epoch denoting the time span between the age of the oldest rocks and formation of Earth), Earth is believed to have undergone heavy meteoric bombardment during this period. It was bombarded by planetoids and other material leftovers from the formation of the solar system. This bombardment, combined with heat from the radioactive breakdown, residual heat, and heat from the pressure of contraction, caused the planet at this stage to be fully molten. Heavier elements sank to the centre while the lighter ones rose to the surface producing Earth's various layers. The bolide activity peaked between 4.0 - 3.85 Gya (Westall et al. 2002). Little is known about geology of the Hadean epoch since most of the remnants of Earth's early curst were mashed and recycled into Earth's interior several times over by plate tectonics since the planet formed. The planet is believed to have cooled quickly with formation of solid crust within 150 million years, and formation of clouds in about 200 million years of Earth's creation. The subsequent rains gave rise to the oceans within 750 million years of Earth's formation, making it a habitable planet for the first time in its history. Until sometime back, the best evidence for the antiquity of the Earth's oceans was from the > 3.7 Gya water-lain sediments of Isua greenstone belt, west Greenland. Now, it is considered that the oceans are probably much older than this. Several new, but indirect, lines of evidence point to the existence of liquid water quite early in Earth's history. Perhaps, the most striking of these is the observation that some carbonaceous chondrite meteorites display hydrothermal alteration meaning thereby that there was water present on their parent asteroid, and liquid water was present even during the process of planetary accretion (Zolensky 2005). Mojzsis et al. (2001) have reported evidence of a possible Hadean ocean as early as 4.3 Gya.

It may not be out of place to mention here that liquid water is the most essential ingredient to trigger beginning of life. Water provides an excellent environment for the formation of complicated carbon-based molecules that could eventually lead to emergence of life. Ehrenfreund et al. (2002) have reported that life on Earth may have emerged during or shortly after the late heavy bombardment phase (4-1 - 3.8 Gya). The precise timing of emergence of life on Earth, however, remains uncertain. More recently, discovery of banded iron formations in northern Quebec (Canada) comprising alternating magnetite and quartz dating back to 4.28 Gya has been claimed to be associated with biological activity (O' Neil et al. 2008). Besides, microscopic analyses of carbon isotope composition of metasediments in Western Australia formed 4.2 Gya have revealed high concentrations of carbon-12 which is typically known to be associated with microbial life (Nemchin et al. 2008).

A prebiotic reducing atmosphere comprising carbon dioxide($CO_2$), methane ($CH_4$), carbon monoxide(CO), ammonia($NH_3$), free hydrogen($H_2$), and water vapour,($H_2O$) if present in Earth's early history, predicts that building blocks of biopolymers - such as amino acids, sugars, purines, and pyrimidines would be formed in abundance (Ehrenfreund et al. 2002). Computer modelling of the Earth's early atmosphere, based on geologic evidence and atmospheric photochemistry, however, suggests in contrast, more neutral conditions (e.g. $H_2O, N_2, CO_2$), thus precluding the formation of significant concentrations of prebiotic organic compounds. According to the present view, the early atmosphere of Earth may have resulted from the mantle degassing of volatiles ($H_2O, N_2, CO,$ and $CO_2$) followed by later accumulation of $O_2$ as product of early life (photosynthesis). The Earth's atmosphere then also perhaps had smattering of $CH_4$ and $NH_3$ (Kasting 2008). It is generally believed that until 2.4 billion years ago, Earth's atmosphere was generally devoid of oxygen. Volcanic activity was intense, and without an ozone layer to hinder its entry, ultraviolet radiation flooded the surface. Thus, the early Earth was just one big chemical evolution experiment (Rollinson 2006). Many scientists now accept the notion that ancient meteorites and comets helped jump-start life on our planet by bringing a significant amount of water, organic molecules and even amino acids to early Earth. Dust particles from comets and meteorites - rich in organic compounds, rained down on early Earth which are believed to have provided an important source of molecules that gave rise to life on our planet. Chyba and Sagan (1992) estimated that >$10^9$ kg of carbon load was delivered to Earth every year from interstellar dust particles, >$10^6$ kg from comets, and by >$10^4$ kg from meteors.

As per the conventional hypothesis, the earliest living cells emerged as a result of chemical evolution on our planet billions of years ago in a process called 'abiogenesis' connoting generation of life on earth from inanimate (non-living) matter. The term is primarily used to refer to the theories about the chemical origin of life, such as from a primordial sea – possibly through a number of intermediate steps. The first indications of life on the Earth come from fossils and carbon inclusion in rocks. The western Australian greenstones, together with similar rocks from Greenland and South Africa are some of the oldest rocks on the Earth. The oldest rocks found on Earth date back to about 4.28 Gyr. The Canadian bedrock along the northeast coast of Hudson Bay has the distinction of being the oldest rock on Earth (O' Neil et al. 2008). As per the palaeontological findings relating to the beginning of life on Earth, 'stromatolites', the oldest microfossils (dome-shaped clumps of bacteria) relating to 11 species of bacteria found in the archaean rocks from western Australia date back to around 3.5 Gyr (Schopf et al. 2002). They are colonial structures formed by photosynthesizing 'cyanobacteria' (blue-green algae) and other unicellular microbes, and are believed to be the 'last universal common ancestor' (LUCA). Cyanobacteria produced oxygen as a by-product of photosynthesis, like today's plants, and played a vital role in the history of our planet in making our atmosphere oxygen rich. This 'LUCA cell' is believed to be the ancestor of all cells and hence all life on Earth. Even older rocks from Greenland, believed to be more than 3.8 billion years old, contain isotopic fingerprints of carbon that could have belonged only to a living being.

In a 1996 paper published in *Nature*, Mojzsis et al. had reported a controversial evidence in *'Nature'* of ancient life dating back to some 3.86 billion years ago found in a rock formation on Akilia island in west Greenland. Scientists look for evidence of life in ancient rocks like those from Akilia island by searching for chemical signatures and isotopic evidence (Mojzsis et al. 1996). The carbon isotope change on the Akilia island rock sample, analyzed with high-resolution microprobe, gave an indication of emergence of life on earth at least 3.86 billion years ago. This was at the end of the period of heavy bombardment of the Earth by comets and meteorites. The researchers found that the ratio of carbon-12 to carbon-13 was 3 % higher than would be expected if life were not present. Since living organisms use the lighter carbon-12, rather than the heavier carbon-13, a lump of carbon that has been processed by a living organism has more carbon-12 atoms than one found in other places in nature. Recently, Manning et al. (2006) at the UCLA Department of Earth and Space Sciences mapped an area on the Akilia island where ancient rocks were earlier discovered by Mojzsis and his teammates to preserve carbon-isotope evidence for life at the time of their formation. Their findings, as reported in the *American Journal of Science*, lend credibility to the fact that these rocks are 3.86 Gyr old and contain traces of ancient life (David and Poole 1999). At the time of the 1996 *Nature* paper, there was no reliable map showing the geology of the area.

Independent studies have provided evidence indicating life may have been present on Earth, fractioning and synthesizing carbon throughout the Hadean epoch. The 'biological fingerprints' from 3.86 Gya were created during the 'Late Heavy Bombardment' period (3.92-3.85 Gya) when the Earth, the moon, and other planets were pummelled with debris that may have harboured complex life. Microbes which took shelter on Earth during the Hadean period may have as their likely source the parent star and its planets (Schoenberg et al. 2002 ; Joseph 2009).

The earliest form of life was a prokaryote - unicellular bacteria possessing a cell membrane and probably a ribosome, but lacking a nucleus or membrane-bound organelles such as mitochondria or chloroplasts, that thrived in aquatic environments, and ruled the Earth in its early history (3 -1.5 Gya) Like all modern cells, it used DNA as its genetic code, RNA for information transfer and protein synthesis, and enzymes for catalyzing reactions. Some scientists believe that instead of a single organism being the last universal common ancestor, there were populations of organisms exchanging genes in lateral gene transfer (Penny and Anthony 1999). For most of the Earth's history, the landscape resembled the present cold Martian desert devoid of eukaryotes - complex multicellular organisms comprising plants and animals including we humans. Eukaryotic cells developed about 1.5 billion years ago in a stable environment rich in oxygen. The rate of evolution of life accelerated at the end of the last 'snowball Earth' about 600 million years ago. About 580 million years ago, the 'Ediacara biota' formed the prelude for the Cambrian explosion. The Cambrian explosion, beginning 542 million years ago, saw rapid appearance of many new species, phyla, and forms over a relatively short period of 5-10 million years. This included emergence of major group of complex animals, as evidenced by fossil record (Butterfield 2001). Prior to that, most organisms were simple - composed of individual cells organized into colonies. The oldest fossils of land fungi and plants date to 480 - 460 million years ago, though molecular evidence suggests the fungi may have colonized the land as early as 1 billion years ago, and the plants 700 million years ago (Heckman 2001). Anatomically, modern humans - Homo Sapiens are believed to have originated somewhere around 200,000 years ago or earlier; the oldest fossil dates back to around 160,000 years ago (Gibbons 2003).

## 3. Spontaneous Generation / Abiogenesis :

Charles Darwin was the first to advocate that "life could have arisen through chemistry in some warm little pond, with all sorts of ammonia and phosphoric salts, light, heat, electricity, etc. present". For much of the 20$^{th}$ century, origin of life research centred around Darwin's hypothesis to elucidate how, without supernatural intervention, spontaneous interaction of the relatively simple molecules dissolved in the lakes or ocean of the prebiotic world could have yielded life's last common ancestor. In 1920s, Alexander I. Oparin in Russia and in J.B.S. Haldane in England revived ideas of spontaneous generation suggesting that the presence of atmospheric oxygen prevented chain of events that would lead to the evolution of life. As early as 1929, Haldane had drawn attention to experiments in which ultraviolet radiation had been seen to encourage build-up of organic compounds from a mixture of water, carbon dioxide and ammonia (Haldane 1947). Oparin argued that a "primeval soup" of organic molecules could be created in an oxygen-less atmosphere, through the action of sunlight (Oparin 1957). Oparin and Haldane thus proposed that the atmosphere of the young Earth, like that of outer (Jovian) planets, was deprived of oxygen or contained very little oxygen($O_2$), and was rich in hydrogen ($H_2$) and other chemical compounds such as methane ($CH_4$) and ammonia ($NH_3$).

Inspired by the ideas of Darwin, Oparin, and Haldane, the duo of Stanley Miller and Harold Urey performed experiments in 1953 (Fig. 1) under simulated conditions resembling those then thought to have existed shortly after the Earth first accreted from the primordial solar nebula, thus heralding era of experimental prebiotic (non-living) chemistry (Miller 1953 ;Urey 1952). Their experiments demonstrated that amino acids and other molecules important to life could be generated from simple compounds assumed to have been present on the primitive Earth. In a self-contained apparatus, Miller and Urey created a reducing atmosphere (devoid of oxygen) that consisted of water vapour, methane, ammonia, and hydrogen above a 'mock ocean' of water. For this, they set up a flask of water to represent the ocean, connected to another flask of gases through which they passed electrical discharge to represent lightning After just two days, they analyzed the contents of the 'mock ocean'. Miller observed that as much as 10-15% of carbon in the system was converted into a relatively small number of identifiable organic compounds, and up to 2% of carbon went into making amino acids of the kinds that serve as constituents of proteins. This last discovery was particularly exciting, as it suggested that the amino acids - basic building components of life would have been abundant on the primitive planet. Miller's experiments produced a cocktail of 22 amino acids and other molecules like purines and pyrimidines, and sugars and lipids associated with living cells (Miller 1953 ; Johnson et al. 2008). According to Miller and Urey, these substances would have been washed into the Earth's early oceans, where they developed into the first living cells. 'Glycine' ($NH_2CH_2COOH$) was the most abundant amino acid resulting from this experiment, and similar other experiments conducted subsequently.

The above experiments showed that some of the basic organic monomers (such as amino acids), which form the polymeric building blocks of a modern life, can be formed spontaneously. Ironically, simple organic molecules are always a long way from becoming a fully functional self-replicating life-form. Moreover, spontaneous formation of complex polymers from abiotically-generated monomers under the conditions presumed in Miller's experiments is not at all a straight forward process. Although it is possible life may have begun on an ancient world whose chemistry was entirely different from Earth, there is nonetheless no evidence to support the belief that life on Earth originated from non-life (Schoenberg et al. 2002 ; Joseph 2009). The first evidence against spontaneous generation came in 1668 from Francesco Redi, who demonstrated that no maggots appeared in meat when flies were prevented from laying eggs. Until Redi's experiments, it was generally held that life had spontaneous beginning. As per the then prevailing popular belief, maggots were spontaneously generated from rotten meat and garbage, and flies came from the chemicals secreted by decaying meat. In 1861, Louis Pasteur too performed a series of experiments wherein he showed that organisms such as bacteria and fungi do not spontaneously appear in sterile, nutrient-rich organic soup.

## 4. RNA World and pre-RNA World Hypotheses :

Biologists, by and large, agree that bacterial cells cannot evolve from non-living chemicals in one step. If life arises from non-living chemicals, it is speculated, there must be some intermediate form of 'pre-cellular life'. Of the various theories of pre-cellular life, in the early stages of evolution of life on Earth, most popular contender today is the 'RNA World' hypothesis (Gilbert 1986 ; Orgel et al. 1968 ; Maddox 1994). It encompasses polymerization of nucleotides into random RNA molecules that might have resulted in self-replicating ribozymes (enzyme). The discovery of ribozymes in the 1980's led to the "RNA World" hypothesis. Seven classes of ribozymes are known to exist in nature. RNA (ribonucleic acid) is a nucleic acid consisting of nucleotide monomers, that act as a messenger between DNA (deoxyribonucleic acid), which is present in all living organisms as the carrier of genetic information, and ribosome that is responsible for making proteins out of amino acids. Single rather than double-stranded, RNA is a chip from original DNA, first hypothesized in late 1950s to act as the first in a chain of intermediates leading from DNA to protein (DNA > RNA > Protein). It is thus the only known macromolecule that can both encode genetic information, and also act as a biocatalyst.

In 1968, Orgel and Francis Crick had proposed that that RNA (Fig. 2) must have been the first genetic molecule. According to them, evolution based on RNA replication preceded the appearance of protein hypothesis. They also held the view that RNA, besides acting as a template, might also act as an enzyme catalyst, and in so doing, catalyzes its own self-replication. Because it can replicate on its own, performing the task of both - DNA and protein, RNA is believed to have been capable of initiating life on its own in the early history of Earth. In the early stages of life's evolution, all the enzymes may have been RNAs, not proteins. The precise events leading to the RNA world however, remain unclear so far. The self-replicating RNA molecules were believed to be common 3.8 billion years ago though the interval in which the biosphere could have been dominated by RNA-based life forms is believed to be less than 100 million years. According to the RNA World hypothesis, descendents of the earliest life form, that relied on RNA for both to carry genetic information and to catalyze biochemical reactions like an enzyme, gradually integrated DNA and proteins into their metabolism.

A few biologists like Senapathy, however, insist that DNA could initiate life on its own (Senapathy 1994). But even the shortest DNA strand needs protein to help it replicate. This is the 'chcken and egg' problem (genes require enzymes; enzymes require genes).Which came first, the chicken or egg? DNA holds the recipe for protein construction. Yet, that information cannot be retrieved or copied without the assistance of proteins. Which large molecule, then, appeared first in getting life kick-started – proteins (the chicken) or DNA (the egg)? A simple solution to the 'chicken-and-egg' riddle, according to Walter Gilbert (1986), is: "one can contemplate an 'RNA World', containing

RNA molecules that serve to catalyze the synthesis of themselves, and the first step of evolution proceeds then by RNA molecules performing the catalytic activities necessary to assemble themselves from a nucleotide soup". In his vision, the first self-replicating RNA that emerged from non-living matter carried out functions now executed by RNA, DNA and proteins. A number of additional clues seemed to support the idea that RNA appeared before proteins and DNA in theevolution of life. Many small molecules called 'cofactors', play an important role in enzyme- catalyzed reactions. These cofactors often carry an attached RNA nucleotide with no obvious functions. These structures have been designated as "molecular fossils", relics that descended from the time when RNA alone, without DNA or proteins, ruled the biochemical world. However, the million dollar question is : how exactly did the first self-replicating RNA arise? At the same time, chemists have been able to synthesize new ribozymes that display a variety of enzyme-like activities (Shapiro 2007).

Another interesting idea is the "PNA" world proposed by Stanley Miller and Orgel (1974). Since starting the RNA World is so difficult, there probably needs to be a pre-RNA world. The PNA (peptide nucleic acid) was hypothesized to be the first prebiotic system capable of self-replication on early Earth (Bohler 1995). The auto–catalytic properties of PNA have not been explored as extensively as for RNA, though PNA's monomers, the diamino acids (also known as diamino carboxylic acid) have been identified in meteorites (Meirerhenrich et al. 2004). Neither ribose nor deoxyribose, or for that matter any other sugars have been found in any prebiotic sources. It was proposed that the earliest life on the Earth may have used PNA as a genetic material due to its extreme robustness, simpler formation, and possible spontaneous polymerization at boiling temperature (Wittung et al. 1994). PNA has bases bound to a peptide-like backbone like diamino acids - a class of amino acids, in contrast to the sugar-phosphate backbone of RNA. This would be pre-cellular life. There is no remnant or trace evidence available whatsoever of the pre-cellular life anywhere today. Its existence is deemed to be entirely conjectural, nor any pre-cellular life has ever been created in a laboratory. Even if the pre-cellular life is presumed to exist in the early epoch of the Earth's existence, there is a problem getting from there to proteins, genes, and cells. The random production of protein does not succeed as an explanation. Although its emergence from non-living matter is hard to conceive, pre-cellular life is believed to have appeared almost spontaneously. Even here, science remains clueless about the timing of spontaneous infusion of life into pre-cellular entity. Researchers are also unclear how even some of the shortest amino acid chains , called peptides, formed prior to the dawn of the living organisms.

Powner et al. (2009) have recently provided some evidence for earthly abiogenesis. They have demonstrated that when the structurally simplest sugar, glycolaldehyde (($CH_2OHCHO$) reacts with cyanamide ($CN_2H_2$),the simplest derivative of cyanide and ammonia, not only the standard nucleobases , but also many undesired compounds were formed. When a third ingredients phosphate is added to the above product, activated pyrimidine ribonucleotides can be formed in short sequence that bypasses ribose and the nucleobases (cytosine and uracil), and in instead proceeds through arabinose amino- oxazoline and anhydronucleoside intermediates. The starting materials for the synthesis–cynamide, glycolaldehyde, and inorganic phosphate are deemed to be plausible prebiotic feedstock molecules (Thaddeus 2006: Sanchez 1966: Pasek 2005: Robertson 2005). Though Powner and colleagues' synthetic sequence yields the pyrimidine nucleotides, it does not explain how purine nucleotides incorporating guanine and adenine could be formed (Szostak 2009). However, based on the works of Proust (1806,1807), the polymerization of hydrogen cyanide (HCN) in aqueous solution has been shown to result in a number of molecules that feature in biological systems. These include purines - adenine, hypoxanthine, guanine, and xanthine (Oro 1960; Sanchez et al. 1967; Roy et al. 2007) and some amino acids (Matthews 1991).

HCN is thus considered to be the precursor molecule of nucleic bases. More recently, large variety of nucleic bases including adenine, cytosine, uracil, and thymine are reported to have been synthesized from formamide ($NH_2COH$) – product of chemical reaction of $H_2O$ and HCN (Saladino et al. 2009). Roy et al. (2007) have proposed blueprint for the possible pathway from HCN to purines under prebiotic conditions (Fig. 3). The production of purine from HCN led to the proposal that purines may have served as the first nitrogeneous bases involved in coding functions in biology (Joyce 1989). Holm et al. (2006) speculate that boron-stabilized ribose, (Ricardo et al.2004) formed in hydrothermal vents via HCN-driven purine synthesis, may have led to the formation of first ribonucleosides. A major challenge to invoking cyanide polymerization for purine synthesis, however, is the availability of HCN itself. While HCN forms readily in gaseous phase, the formation of HCN at biologically relevant locales is more problematic as the areas where molecular organization relevant to origin of life takes place are generally aqueous, not gaseous environment (Matthews 1991 ; Minard et al. 1998).

Christian de Duve(1995) posited a "Thioester World" which is believed to have preceded the RNA world. The thioester world represents a hypothetical early stage in the development of life that could have provided the energetic and catalytic framework of the protometabolic set of primitive chemical reactions that led from the first building blocks of life to the RNA world, and subsequently sustained the RNA world until metabolism took over. A thioester forms when a 'thiol' or sulfhydral (whose general form is written as an organic group, R, bonded with sulfur and hydrogen, hence R-SH) joins a carboxylic acid (R'-COOH). A molecule of water is released in the process, and what remains is a thioester: R-S-CO-R'. The thioester bond is a high-energy bond, deemed to be equivalent to the phosphate bonds in adenosine triphosphate (ATP), which is the main supplier of energy in all living organisms. Constituents of of thioester bond are believed to be available in prebiotic soup in plenty. Amino acids and carboxylic acids are the most conspicuous substances found in meteorites, and thiols that are expected to arise readily in the kind of volcanic setting, rich in hydrogen sulfide ($H_2S$). They are likely to have been abundantly available on the prebiotic Earth. Thioesters, thus, are believed to have played the role of ATP in a thioester world initially devoid of ATP (De Duve 1995). The thioester world hypothesis, however, is highly speculative, being supported largely by the need of congruence between protometabolism and metabolism, though experiments performed in the past have provided some evidence in support of this hypothesis. In a laboratory experiment, researchers have obtained under plausible conditions the three molecules – cysteamine, b-alanine and pantonic acid - that make up a neutral substance known as 'pantetheine', which is the most important biological thiol, a catalytic participant in a vast majority of the reactions involving thioester bonds (De Duve 1995).

Networks of inorganic compartments formed, primarily, of iron sulfide and existing in the vicinity of hydrothermal vents could be veritable reactors for RNA synthesis (Koonin 2007). At high concentrations of nucleotides, the equilibrium, however, shifts from hydrolysis to polymerization, and the rate of reaction could be enhanced through the combined effects of thermophoresis involving thermal gradients and inorganic catalysis (Ogasawara et al. 2000). Thus, the putative reactors at hydrothermal vents could provide the exact substrate that is needed for the evolution of a primordial RNA world comprising of a dense population of versatile RNA molecules, some of which could possess capability as a catalyst. This system might facilitate emergence of ribozyme RNA replicases and other ribozyme activities that are required for the onset of an evolving RNA world Gesteland et al. 2006).

Although Joyce (2002) recognizes 'RNA World' as the leading account of the origin of life from non-life. He also acknowledges many of the problems inherent in the 'RNA World' hypothesis. The foremost obstacle to understanding to origin of RNA-based life is identifying a plausible mechanism for overcoming the clutter wrought by prebiotic chemistry. Other notable problems with the hypothesis relate to the instability of RNA when exposed to ultraviolet light, the difficulty of activating and ligating nucleotides and lack of available phosphate in solution required to constitute the backbone, and the instability of the base cytosine which is prone to hydrolysis. RNA is chemically fragile and difficult to synthesize abiotically. The known range of its catalytic activities is rather narrow, and the origin of an RNA synthesis apparatus is unclear. To solve some of these problems, Joyce suggests, 'some other genetic system proceeded' RNA, just as it preceded DNA and protein'. He further adds that RNA-based functions for which there is no evidence in biology, such as nucleotide synthesis and RNA polymerization, are assumed to have existed in the RNA World on first principles.Ironically, the assumption is not supported by available historical evidence. Serious threat to validity of the RNA world hypothesis also arises from the fact that an RNA world consisting of RNA molecules is unlikely to occur because the biologically important interactions are not effective for the bare RNA molecules at high temperatures prevailing in the early epoch of the Earth (Kawamura 2004).

Some life science experts believe life had its first foothold in a hot spring on the outgassing early Earth (Hartman 1998). The first organisms were self-replicating iron-rich clays which fixed carbon dioxiode into oxalic and other dicarboxylic acids. This system of replicating clays and their metabolic phenotype then evolved into sulfide-rich region of the hot spring acquiring the ability to fix nitrogen. Finally, phosphate was incorporated into the evolving system which allowed synthesis of purine and pyrimidine bases. Subsequent polymerization of the amino acid thioesters into polypeptides preceded the directed polymerization of amino acid esters by polynucleotides. The origin and evolution of the genetic code is thus a late development. Role of clay is eventually taken over by RNA.

Based on laboratory experiments, some researchers have reported that it would have been possible for genetic molecules similar to DNA or its close relative, RNA to form spontaneously. The earliest forms of life are conjectured to have been simple membranes made of fatty acids known to form spontaneously that enveloped water and the self-replicating genetic molecules. The genetic material would encode the traits that are alive today. Fortuitous mutations appearing at random in the copying process, would then propel evolution, enabling theses early cells to adapt to their environment,to compete with one another, and eventually turn into life-forms we know (Ricardo and Szostak 2009).

## 5. Extremophiles :

Extremophiles (Mondigan and Marrs 1997; Fredrickson and Onstott 1996; Brock 1978) are usually unicellular microbes (bacteria and archaea) that can survive in the harshest of environment on the planet – considered extremely inhospitable for habitation by humans and other creatures. Many of them are evolutionary relics called "archaea", believed to be among the first homesteaders on the Earth 3.8 billion years ago. They are presumably the first version of life on our planet when its atmosphere was devoid of oxygen, and comprised largely of ammonia, methane, water vapour and carbon dioxide. They are microorganisms similar to bacteria in size and simplicity of structure, but continue as an ancient group intermediate between bacteria and eukaryotes. Heat–loving microbes, or "thermophiles", are among the best studied of the extremophiles that can reproduce or grow readily in temperatures exceeding 45ºC, and some of them, referred to as "hyperthermophiles", are capable of thriving in temperatures as high as 110ºC – more than boiling water temperature. These microbes are known to have colonized a variety of extremely hot places such as hot springs, deep oceanic vents, and the deep subsurface. In New Zealand, microbes have been found to flourish in pools of temperature up to 101ºC (Postgate 1994). In the hot, acidic springs of Yellowstone National Park (USA), microbes have been found to be thriving in temperatures of upto 95ºC. Communities of hyperthermophiles have been recovered from geothermally heated rocks 2,500 - 3,500 metres beneath the surface of the Earth, and aside 400ºC rock chimneys at depths of 4,000 metres, at the bottom of the ocean where pressures are up to 62 Mpa (~612 atmosphere) (Boone et al. 1995). "Baccilus infernus" is known to thrive at depths of 2,700 metres where the weight and pressure is 300 times of the surface, and where temperatures may exceed 117ºC (Boone et al. 1995; Setter 2002). Explorations to various hot springs and deep-sea hydrothermal vents have led to identification of more than 50 species of 'hyperthermophile'.

The majority of the hyperthermophiles found near the black chimneys (vents) are anaerobes, thriving in 100ºC boiling water, that can survive without oxygen, liberating chemicals and minerals for energy (Boone et al. 1995). Hyperthermophiles, living in and around the dark hydrothermal vents along the mid-oceanic ridges, harness chemical energy from hydrogen sulphide ($H_2S$) and other molecules that billow out of the sea-floor at temperatures as high as 121ºC (Brock 1978). The ability of hyperthermophiles to tap such geothermal energy raises interesting possibilities for other worlds like Jupiter's moon, Europa which probably harbours liquid water beneath its icy surfaces ( Is there life on Europa? 2004). Europa is squeezed and stretched by gravitational forces from Jupitar and other Galilean satellites. Tidal friction heats the interior of Europa possibly enough to maintain the solar system's biggest ocean. It is guessed that similar hydrothermal vents in Europa's dark sea fuel ecosystems like those found on Earth. Based on these speculations, existence of water oceans even on other moons in the solar system like Jupiter's other satellites – Callisto and Ganymede, Enceladus (Saturn's moon) and Triton (Neptune's moon) with the possibility of life thriving within their interiors cannot be ruled out. Radar observations made by the Cassini spacecraft regarding Titan's rotation and shifts in the location of surface features are suggestive of a vast ocean of water and ammonia lurking deep beneath its surface (Lorenz 2008). Microbes (methane-oxidizing archaea) found thriving at a record depth of 1.62 km beneath the Atlantic seabed at simmering temperature range of 60-100ºC give rise to the possibility that life might evolve underground on other planets and their satellites as well (Roussel 2008). The living prokaryotic cells in the searing hot marine sediments have ranged in age from 46 to 111 million years.

Some thermophiles live at remarkable depths within the Earth's crust. In South Africa, samples of rock from gold mines at 2.8 km depth below the surface show the presence of thriving thermophiles that derive all its energy from the decay of radioactive rocks rather than from sunlight (Lin et al. 2006). The self-sustaining bacterial community represents the first group of microbes known to depend exclusively on geologically produced hydrogen and sulphur compounds for nourishment, and lives in conditions similar to those of early Earth. The subterranean world, according to Lin and his colleagues. whose findings were reported in a 2006 issue of *Science*, is a lightless pool of hot and pressurized salt water that stinks of sulfur and noxious gases human would find unbearable. The microbes discovered in the South African mines are related to the 'firmicutes' division of microbes that exist near the hydrothermal vents, and appear to have survived for tens of million years without any nutrients derived from photosynthesis. The bacteria's rocky living space is a metamorphosed basalt that is 2.7 billion years old. How the surface-related 'firmicutes' managed to colonize in an area so deep within Earth's crust is a mystery. Some surface 'firmicutes' are known to consume sulfate and hydrogen as a way to get energy for growth. As per the DNA analysis of the bacterial genes, the subsurface 'firmicutes' were removed from contact with their surface cousin anywhere from 3 million to 25 million years ago (Lin et al. 2006). The extreme conditions under which the bacteria live bear a resemblance to those of the early Earth, potentially offering insight into the nature of organisms that lived long before our planet had an oxygen-filled atmosphere.

## 6.   Thermophiles and Deep-sea-origin of life :

Hyperthermophiles, having the capability to thrive at temperatures between 80-121ºC such as those found in hydrothermal systems, are deemed to be our closet link to the very first organisms to have evolved on the Earth(Brock 1978). At the beginning of life, the Earth was a much hotter planet due to increased greenhouse effect of a carbon dioxide - rich atmosphere. This early atmosphere also did not contain oxygen until the 'Great Oxidation Event' 2.3 - 2.4 billion years ago. Because life started about 3.8 billion years ago, it was probably exclusively anaerobic for at least first 1.5 billion years. Unique heat resistance and anaerobic nature of many hyperthermophiles could be the traits of earliest organisms on the Earth as well.  Astrobiologists are increasingly becoming convinced that life on Earth itself might have started in the sulfurous cauldron around hydrothermal vents.  Indeed, many of the primordial molecules needed to jumpstart life could have been found in the subsurface out of the  interaction of rock and the circulating hot water driven by the hydrothermal systems.

The "iron-sulfur world", first hypothesized by Gunter Wächtershäuser in 1980s, identifies the "last universal common ancestor (LUCA) " inside a black smoker at ocean - floor, rather than assuming free-living form of LUCA. The 1977 discovery of an entire ecosystem inside a black smoker (Fig. 4 and 5) at the ocean floor, using an US submarine, that did not rely on photosynthesis for its survival, forced scientists to have a re-look at the prevailing view about origin of life (Shrope and Pickrell 2006).  Hydrothermal vents in the deep ocean typically form along the mid-ocean ridges, such as 'East Pacific Rise' and the 'Mid-Atlantic Ridge'. In contrast to the classical Miller experiments, which depend on the external sources of energy such as simulated lightning or ultraviolet irradiation, Wächtershäuser argued that life was more likely to have arisen in the exotic conditions near volcanic vents, driven by metal catalysts (Huber and Wächtershäuser 1998). Wächtershäuser systems come with a built-in source of energy - sulfides of iron (FeS) and other minerals (e.g.pyrite) which help in evolving autocatalytic set of self-replicating, metabolically active entities that would predate life forms known today.

Geologists have discovered 1.43 billion year-old fossils of deep-sea microbes, providing more evidence that life may have originated on the bottom of the ocean.  The ancient black smoker chimneys, unearthed in a Chinese mine, are nearly identical to archaea and bacteria - harbouring structures found on sea-beds (Li and Kusky 2007). These chimneys develop at submerged openings in the Earth's crust that spew out mineral-rich water as hot as 400ºC compared to almost freezing temperature (~2ºC) for the surrounding deep ocean waters. 'Arqueobacteria',  that do not depend on sunlight or oxygen move into fragile chimneys that grow around the vents, and  feed on the dissolved minerals particularly hydrogen sulphide to produce organic material through the process of chemosynthesis. Interestingly, all biological molecules are normally known to be destroyed at a temperature of 150ºC. Prebiotic enzyme-like assemblies, however, could have facilitated the accumulation of RNA molecules at the hydrothermal vent temperatures (Kawamura 2004).The above instance appears to be reinforcing the views of Wächtershäuser and those of William Martin and Michael Russel(2002) that the first cellular life forms may have evolved originated on the bottom of the ocean inside black smokers at seafloor spreading zones in the deep sea .

The deep-sea vent theory posits that life may have begun at submarine hydrothermal vents, where hydrogen-rich fluids emerge from below the sea-floor and interface with carbon dioxide-rich ocean water. Sustained chemical energy on such systems is derived from redox reactions, in which electron donors, such as molecular hydrogen, react with acceptors, such as carbon dioxide. The study of Maher and Stevenson (1988) suggests that if deep marine hydrothermal setting provides a suitable site for origin of life, abiogenesis could have happened on Earth as early as 4.0 - 4.2 Gya, whereas if it occurred at the surface of the Earth, abiogenesis could only have occurred between 3.7 - 4.0 Gya.

## 7. Other Extremophiles :

Hardy microbes (Brack et al. 1994) have been found to flourish in other extreme environments such as extreme acidic, alkaline, and saline as well. 'Acidophiles' thrive in caustic environment with pH level at or below 3, and 'alkaliphiles' favour habitat in an alkaline environment with pH levels of 9 or above. Most natural environments on the Earth are essentially neutral having pH values between 5 and 9. Highly acidic environments can result naturally from geochemical activities such as the production of sulfurous gases in hydrothermal vents and some hot springs. Acidophiles are also found in the debris leftover from coal mining. Alkaliphiles live in soils laden with carbonate and in so called soda lakes as those found on Egypt, the Rift Valley of Africa and the Western U.S. 'Halophiles' make home in intensely saline environments, especially natural salt lakes like the Great Salt Lake (USA). A new species of living bacteria , "spirocheta americana" has been found flourishing deep inside California's Mono Lake's salty-alkaline mud where no oxygen could reach ( NASA News Release 2003).

Recent experiments suggest that if bacteria were somehow sheltered from the radiation of space, perhaps inside a thick meteoroid, they could survive in dormant state for millions of years. Dormant bacterial spores are reported to have been revived in 30 millions old amber. A new species of bacterium , "herminiimonas glacei", found 3 km deep in Greenland's glacial ice, is reported to have survived 120,000 years of freezing (Loveland-Curtze et al. 2009). Freeze-dried microorganisms have been recovered from permafrost and from depths of subglacial Lake Vostok (Antarctica), and cultured in the laboratory after lying dormant for 10-20 million years. Semi-dormant bacteria found in ice-cores over a mile beneath the Antarctica lends credibility to the idea that the components of life might survive on the surface of icy comets. 'Psychophiles' are the other cold-loving microbes that thrive in the most frigid places like the Arctic and Antarctica that remain frozen almost throughout the year. Microorganisms have been recovered from cores 400 metres deep in the Canadian Arctic (Gilichinsky 2002). According to Gilichinsky, " the permafrost community has overcome the combined action of extremely cold temperature, desiccation, and starvation, and life might be preserved in permafrost condition for billion years". Bacteria have also been recovered in salt crystals from a New Mexico salt mine dated at 250 million years (Vreeland et al. 2000). Beside, 'living bacteria' have been discovered and isolated from the Middle Devonian, the Silurian, and the Precambrian periods making some of them over 600 million years in age (Dombrowski 1963).Thus, there is all possibility that bacterial spores could survive for a very prolonged period of 250 - 600 million years, which is more than enough time for them to take up residence on planets made of the debris from supernova explosion.

Then, there are bacteria that do not rely on photosynthesis for energy at all. In particular, 'endolith' bacteria using chemosynthesis have been found to survive in microscopic spaces within rocks such as pores between aggregate grains, and in subterranean lakes. "Hypolith" bacteria live inside rocks in cold deserts. Of all different strains of bacteria on Earth, those in genus 'Deinococcus' are really a hardy bunch. They are extremely resistant to ionizing radiation as well as nuclear radiation. 'Deinococcus geothermalis' can sustain the harshest environment on the planet- its favoured habitat includes nuclear plants. 'Deinococcus radiourans', that thrives within the cores of nuclear reactors, is capable of withstanding 500 times radiation that will easily kill a human - with no loss of vitality. Such bacteria perform the amazing effect of using an enzyme to repair DNA damage, in cases where it is estimated that the DNA experienced as many as a million breaks in its helical structure (Secker et al. 1994).

Many species of microbe have evolved extraordinary ability to survive a violent hypervelocity impact, and extreme acceleration and ejection into space including extreme shock pressure of 100 Gpa; the frigid temperatures and vacuum of an intersteller environment; the descent through the atmosphere and landing onto the surface of a planet (Burchell et al. 2004 : Burchella et al. 2001 : Horneck et al. 2001 : Mastrappa et al. 2001).

## 8. Extremophiles and Astrobiology :

Astrobiology is the field concerned with forming theories about the distribution, nature and future of life in the Universe. Astrobiologists are particularly interested in studying extremophiles, as many organisms of this type are capable of surviving in environments similar to those known to exist on other planets. Great distribution and diversity of microbial extremophiles on Earth have profoundly increased the probability that life may also exist elsewhere in the cosmos. For example, Mars may have regions in its deep subsurface permafrost that could harbour 'Endolith' communities. From a microbial point of view, environmental conditions on Earth and Mars being similar, early Earth and early life represent an excellent analogue for the study of potential Martian life (Westall et al. 2002).The subterranean ocean of Jupiter's Europa may harbour life at the hypothesized hydrothermal vents at the ocean floor. Another hyperthermophile that lives in deep sea chimneys - the methane-producing archaean, 'methanopyrus' is being studied to help understand how the world's earliest cells survived the harsh environs of the primitive Earth. Even acidic clouds of Venus too are suspected of harbouring life. The mysterious dark patches in the Venusian atmosphere may be related to the vast communities of hardy microbes comparable to the terrestrial extremophiles (Stuart 2002). Moreover, bacteria are space hardy, and have the capability to sustain extremes of temperature and pressure. Biological catalysts are able to convert hydrogen and carbon dioxide or carbon monoxide into methane with ease. Class of bacteria known as 'methanogens' is capable of achieving this fact. Given the fact that methane is found in large quantity in the atmospheres of the four Jovian planets – Jupiter, Saturn, Uranus, and Neptune, presence of methanogens cannot be ruled out. Also, Titan, the largest satellite of Saturn, where a methane lake of the size of Caspian sea has been spotted, presence of Methanogens there too could not be ruled out (Stofan et al. 2007).

Psychrophilic and psychrotrophic (cold-loving) microbes that inhabit permanently frozen regions of permafrost, polar ice-sheets, and glaciers, and deep ocean and deep-sea sediments on Earth provide analogues for microbial life that might inhabit ice sheets and permafrost of Mars, comets, or the ice/water interfaces or sediments deep beneath the icy crust of Europa, Callisto, or Gnymede (Pikuta and Hoover 2003).

## 9. Panspermia :

Panspermia provides alternative to earthly 'abiogenesis' (generation of life from non-living matter). It is founded on the belief that "life comes from life", and is currently regarded as a viable hypothesis for the beginning of life on our planet. It hypothesizes that the primitive life on Earth may have originally formed extraterrestrially. As per this concept, seeds of life are ubiquitous, and they may have delivered life to the Earth as well as to the other habitable bodies in the Universe. Seeds of life, according to panspermia, travel across galaxies as 'life spores' protected in comets and asteroids from ultraviolet radiation. These life spores seeded the planets in the solar system and other cosmic entities in the Universe. Bacterial spores are known to survive ultraviolet exposure in satellite experiments lending support to the panspermic view for origin of life. A thin layer of graphitized carbon around individual bacteria, only 0.02 µm thick, effectively blocks damaging effects of ultraviolet light (Mileikowsky et al. 2000 ; Wickramasinghe, N.C. & Wickramasinghe, J.T. 2003). Extraordinary capability of some extremophiles and bacterial spores to survive in extreme environments on Earth also favours the perception that simple life-forms may have originated between the stars or been capable of surviving long interstellar journeys.

There has been reference to cosmic nature of life in the ancient Vedic and Buddhist scriptures. Prominent Greek philosopher, Anaxagoras who lived around 500 BC, too believed in the concept of universal spreading of life. In modern times, Svante Arrhenius, a Swedish Nobel laureate promoted the concept of panspermia in the right earnest at the dawn of the twentieth century (1903). Arrhenius held the view that microbe spores were propelled through space by radiation emitted by stars, and they were possibly the seeds of life on the primitive Earth. He recognized the importance of the solar radiation pressure in that its pressure on small particles would be greater than the solar gravity (28 times Earth's gravity), and calculated that small particles with diameter 0.16 ɪm are most susceptible to solar radiation pressure, and such particles would easily be expelled from the solar system. The spores of bacteria have a diameter slightly greater than this value (the spore of 'Bacillus subtilis' has a diameter of 0.5 µm). Arrhenius then argued that if the spores or smaller organisms could be released from the terrestrial gravity, the solar radiation pressure would expel them out of the solar system, and if this process continues, the whole galaxy would be populated by organisms not too different from those on Earth (Yabushita et al. 1986).

Life, as per the panspermic view, is believed to have been introduced to Earth some 4 billion years ago, with an ongoing incidence of microorganisms arriving on the Earth from the space hitchhiking on comets and meteorites, that continues till today. The role of comets, carbonaceous meteorites, interstellar dust and asteroids in the delivery of water, organics and prebiotic chemicals to Earth during the Hadean epoch of heavy bombardment has become widely recognized in recent years (Hoover 2002). Carbonaceous chondrites contain different classes of molecules, many of which are common biochemical compounds, such as amino acids and their precursors, nucleic acid bases, and sugar-like molecules. Interesting prediction of panspermia is that life throughout the universe is derived from the same ancestral stock. Fred Hoyle and Chandra Wickramasinghe have been prominent proponents of this hypothesis after Arrheniuus, who believed that life forms continue to enter the Earth's atmosphere, and they may also be responsible for epidemic outbreaks, new diseases, and the genetic novelty necessary for microevolution (Hoyle and Wickramasinghe 1979). They firmly supported the view that a life-bearing comet - rich in organic content struck the Earth about 4 billion years ago, depositing its cargo of primitive cells – the forerunner of all the life today. The comets in the solar system collectively contain ~$10^{30}$ gm of carbonaceous matter which could be returned in the form of bacterial dust into interstellar space. The dominant source of exogenous organics to early Earth is believed to be interplanetary dust particles, which appear to be ~10% organic by mass, and decelerate gently in the atmosphere , and so can deliver their organics intact. Particles < 1μm across do not burn up in the atmosphere, but fall gently to the ground. As per one estimate, the Earth is sweeping 3,000 tonnes of interplanetary dust per year, providing about 300 tonnes of organic materials (Chyba and Sagan 1992).

Hoyle and Wickramasinghe strongly advocated that dust grains in the interstellar medium had a complex organic composition, and their inclusion within comets led to transformation of pre-biotic matter into primitive bacterial cells. They supported the argument that dust grains in the interstellar clouds contain bacteria such as Escherichia–coli (commonly known as E-coli) which survive in comets. Their argument was based mainly on excellent matching of the UV extinction curves with the laboratory opacity curves for terrestrial spores forming bacteria. They also obtained agreement of the observed infrared fluxes with the predicted curves for bacteria and bacteria-silica mixture (Wickramasinghe 1989). The first identification of organic polymers in interstellar grains was made by Wickramasinghe in 1974, and the first association of a biopolymer with interstellar dust was discovered in 1977 by Hoyle and Wickramasinghe. The microorganism model of interstellar grains has since been investigated through spectroscopy - from infrared (IR) to visible to ultraviolet (UV) bands with bacterial extinction matching interstellar extinction to a reasonably good degree of precision. The first laboratory spectral signature characteristic of freeze-dried bacteria was identified by Sirwan Al-Mufti in 1982 during analysis of infrared absorption by dust in the waveband of 2.9-3.9 μm for the galactic centre infrared source GC-IRS7 (Hoyle et al. 1982). The spectrum of this IR source revealed a highly detailed absorption profile over the above waveband, indicative of combined CH, OH, and NH stretching modes. However, a laboratory spectrum of the dessicated E-coli still provided an exceedingly close point-by-point match (Fig. 6) to the astronomical data of GC-IRS7(a late-type supergiant) over the entire 2 - 4μm waveband (Wickramasinghe 2004) lending support to the microbial model for interstellar grains. Moreover, Hoyle et al. (1985) claimed that the well-known 2200 Å peak in the interstellar extinction curve, generally attributed to small graphite grains with a smattering of PAH molecules, could be better interpreted in terms of biological cells. Their reasoning for the above interpretation was based on observation of a 2200 Å peak in the spectra of E-coli and yeast.

Interstellar and cometary dusts have since been found to have spectroscopic properties consistent with widespread occurrence of microbial material. Studies of the comet Halley in 1986 revealed infrared signatures indicative of bacterial content. Other comets too have shown similar effects in the past, not only over 2.9 - 3.5 µm waveband, but also at longer infrared wavelengths. The pre- existing viable bacterial cells, derived from interstellar space, are believed to have been included in comets and meteorites in the primitive solar system 4.5 billion years ago (Hoyle and Wickramasinghe 1977). The cometary collisions with the Earth may themselves have injected life onto the Earth on several occasions during the Hadean Epoch – some 4.3 - 3.8 billion years ago. Comets (about 100 billion), residing in the Oort cloud within the outer reaches of the solar system with profuse of organic materials, are believed to be the ideal sites for origin of simple living organisms. The young Earth could have also received more complex molecules with enzymatic functions, molecules that were pre-biotic but part of a system that was already well on its way to biology. After landing in a suitable habitat on our planet, these molecules could have continued their evolution to living cells. Thus, life could have roots both on Earth and in space (Warmflash and Weiss 2005).

Some skeptics, however, have reservations about the validity of microbial model for interstellar grains. Wada et al. (1992) have reported that their spectroscopic studies on biological cells and organic extracts from carbonaceous compounds have failed to identify the well-known 2200 Å interstellar extinction peak with the organic material. Then, Zagury (2002) has reported his review of a series of observations in *New Astronomy* that do not agree with the standard interpretation of the interstellar extinction curve. His study suggests that a true extinction curve ought to be a straight line from near IR to the far UV. If so, validity of interstellar grain models would be very much debatable. Moreover, claim for close fittings of the UV extinction curves with the laboratory opacity curves for terrestrial spores forming bacteria as also good agreement of the observed infrared fluxes with the predicted curves for bacteria and bacterium-silica mixture have not been considered by some scientists as being above suspicion. The controversy arises from the fact that bacteria grain spectra are taken in solid state, which generally do not have any sharply localized salient features like the well-defined rotation or rotation-vibration in the gas phase molecular spectra (Moore and Donn 1982).

Stellar nucleosynthesis of heavy elements such as carbon allowed the formatiom of organic molecules in space, which appear to be widespread in our Galaxy. Discovery of 'glycine' ($CH_2NH_2COOH$)-simplest amino acid (building block of protein) and the simplest sugar-glycolaldehyde ($CH_2OHCHO$) detected in millimeter-wave rotational transitional emission from Sagittarius B2, a dense star-forming cloud of interstellar gas at the very heart of the Milky Way, bolsters the view that interstellar medium may have played a pivotal role in the prebiotic chemistry of Earth(Sorrell 2001). Researchers, using radio astronomy, claim that the spectral fingerprint of glycine in the frequency range of 90-265 GHz is the first step in establishing the critical link between amino acids in space and the emergence of life in the solar system or elsewhere in the Milky Way Galaxy (Kuan et al. 2003). The ubiquitous presence of organic molecules in the interstellar clouds, comets, and asteroids, and the evidence of extraterrestrial amino acids in carbonaceous meteorites lend support to a cosmic perspective on the origin of life. Hoyle and Wickramsinghe related large scale presence of assortment of PAHs in galactic and extragalactic regions to disruption of bio-grains in the presence of UV flux (Wickramasinghe 1989, 1995). According to them, production of the organic matter is due to biological processes wherein bacteria accept inorganic matter as nutrient and grow exponentially analogous to colony formation by the terrestrial bacteria. Observations at infrared, radio, millimeter and sub-millimeter frequencies have revealed presence of over 150 different chemical compounds including several organic compounds with C, H, O, and N as major constituents in the interstellar clouds, circumstellar envelopes, and comets since 1965 (Lammer et al. 2009). They range in complexity from the simplest diatomic ($H_2$) through familiar ones like methane($CH_4$), formic acid (HCOOH), methanol ($CH_3OH$), ethanol ($CH_3CH_2OH$), hydrogen cyanide (HCN), and nitrous oxide or laughing gas ($N_2O$), to thermally hardy compounds such as polycyclic aromatic hydrocarbons (PAHs). Fig. 7 provides near infrared spectrum of ethanol.

In a balloon experiment performed by Narlikar et al. in January 2001, air samples at heights ranging from 20 to 41 km over Hyderabad (India) – above the local tropopause (16 km) where mixing from the lower atmosphere is unexpected, a vast amount of viable but not-culturable microorganisms were discovered along with three species of microorganisms that very much resembled their known terrestrial counterparts namely, cocus (spherical bacterium), bacillus(rods), and a fungus identified as 'Engyodontium albus' lending credence to the concept of panspermia ( Narlikar et al. 2003 ; Wainright et al. 2002). The daily input of such biological material is provisionally estimated to be in the range of one-third to one tonne over the entire planet (Wickramasinghe 2004). Distinct species of over 1,800 different types of bacreia and other microbes thrive and flourish within the troposphere (Brodie et al. 2007). Moreover, microorganisms and spores have been recovered from even at a height of 130,000 ft.(Soffen 1965).

There are also reports of fossilized microorganisms being found inside a 4.6 billion year – old Murchison meteorite (Kvenvolden et al. 1970) that crashed to the Earth (Australia) in 1969. The organisms have been identified as bacteria capable of surviving in the extreme environments that lends credibility to the belief that life could have first evolved elsewhere in deep space, before a meteorite or comet seeded the Earth. Over the past 20 years , 30 meteorites have been found on the Earth that originally came from the Martian crust, based on the composition of gases trapped within some of the rocks. In a disputed claim, a meteorite originating from Mars known as 'ALH84001' was shown in 1996 to contain microscopic structures resembling terrestrial microfossils and a variety of organic molecules, including PAHs (McKay et al. 1996). In the meanwhile, biologists have detected organisms durable enough to survive journey from Mars to Earth, inside such meteorites. Researchers, using the world's largest radio telescope - the Arecibo Observatory in Puerto Rico, have detected an amino acid precursor, 'methanimine' in the far-flung galaxy Arp 220 - some 250 million light years away from Earth, which provides fresh evidence that life has the potential to evolve throughout the Universe. Methanimine can form the simplest amino acid, glycine, when it reacts with either hydrogen cyanide and then water, or formic acid (Salter 2008). Earlier, evidence of formaldehyde, ammonia, hydrogen cyanide, and formic acid was found in the star-forming region. In a controversial claim, microbial fossils have been identified in 15 carbonaceous chondrites (Hoover 2006). Amino acids and nucleobases for DNA and RNA, including adenine, guanine, alanine, and isovaline have also been discovered in interiors of the Orgeuil meteorite. Carbon isotopic measurements demonstrated these acids were extraterrestrial in nature, and originated in an environment with water and a high concentration of organic carbon (Hayatsu 1964; Folsome et al. 1971; Lawless et al. 1972).

## 10.  Concluding Remarks :

 Given the extraordinary potential of extremophiles to survive in highly inhospitable environments on the Earth, possibility of life on extraterrestrial bodies like Mars, Europa, Titan, Enceladus, Ganymede, Callisto, and  Triton in the solar system cannot be ruled out. Also, study of  the planetary geology reveals that our solar system could  have many worlds with liquid water, the essential ingredient for life. Recent data  from NASA's Mars Exploration  Rovers also corroborate the  speculation that water has at least intermittently flowed on the red planet in the distant past (Warmflash and Weiss 2005). Moreover, studies based on data from NASA's Mars Reconnaissance Orbiter have  revealed that the red planet once hosted vast lakes, flowing rivers, and a variety of other wet environments that had the potential to support life. It is thus not unreasonable to hypothesize that life existed on Mars long ago, and perhaps continues there even today. Permafrost environments on Mars may also help harbour life in the light of the fact the ancient bacteria on Earth are capable to spring back to life after being in  state of prolonged hibernation  for nearly half a million years in harsh and frozen conditions.  Besides, analyzing data gathered by Cassini spacecraft, scientists  have recently  confirmed presence  of heavy negative ions about 10,000 times the mass of hydrogen in the upper regions of Titan's atmosphere which is devoid of oxygen and comprises mainly of nitrogen and methane.  These particles may act as building blocks for complicated organic molecules - the harbinger for earliest form of life in Titan's atmosphere (Richardsons et al. 2007)**.** Moreover, Cassini's radar mapping reveals that Titan is just covered in carbon-bearing material. It is a giant factory of organic chemicals with several hundred lakes and sea of hydrocarbons (methane and ethane). Life may have even got a foothold on the torrid Venus.  Though the Venusian  surface is  too hot (~ 480°C) and under too much atmospheric pressure (90 bars) to be habitable, the planet could still conceivably support sulfur-based microbial life high in its atmosphere as do sulphur-eating 'chemotrophs' on the Earth.

Keeping in view the great surviving capabilities of  extremophiles on the Earth, possibility of traces of life forms being found on billions of unexplored planetary bodies outside our solar system cannot be negated as well. Out of  721 exoplanets detected  beyond  our  solar  system  till  January 2012, the  extrasolar  planet (Jupiter like gaseous planet)  named 'HD209458B' located  at some 150 light years from the Earth in the constellation of Pegasus is believed to be harbouring water vapour in its atmosphere giving rise to speculation of the presence of life-supporting microbes in its atmosphere based on the analysis of the infrared spectrum in the range of 7.5 - 13.2 µ m (Maa et al. 2007). The planetary habitability chart (Fig.8) provides clue  where life might exist on extrasolar planets based on  study of our solar system and life on Earth. Recent computer simulations of  the  known extrasolar planetary systems suggest about half of the hitherto known exoplanets could harbour Earth-like world raising possibility of traces of life being found thereon. The author describes at length the possibility of finding traces of life on the habitable planets and moons within the  solar system  and beyond in his  article titled 'Searching for Life on Habitable Planets and Moons' published  in *Journal of Cosmology* ( Lal 2010).

Despite strong possibility of existence of a large number of extraterrestrial life-systems in the Universe, there remain nagging uncertainties in regard to the timing of commencement of the process of evolution of life on the primitive Earth and elsewhere in the Universe. Though panspermia provides satisfactory explanation to the origin of life on Earth and elsewhere in the Universe, it however, fails to address the long-standing riddle as to when and where precisely life originated first in the Universe, nor does it provide any clue about how transformation from pre-biotic matter into primitive bacterial cells was brought about.  Evidence from comets and meteorites as also from the  experiments that simulate the conditions on the early Earth suggests that probably a combination of terrestrial and extraterrestrial factors were responsible for kick-starting the process of transforming pre-biotic organic compounds into entities that we call 'life' on Earth.

There is growing evidence to support the view that emergence of catalytic RNA was a crucial early step in the evolution of life on Earth. How that RNA came into being, however, remains unknown so far. Moreover, the "RNA World" hypothesis does not seem to provide satisfactory explanation to the initiation of mechanism of 'self-replication' in organisms in the early history of Earth, which is so crucial to the understanding of the process of evolution of life on our planet and other habitable bodies in the Universe. We do not yet understand the steps leading from abiotic early Earth to the RNA World( Orgel 1995). Experiments involving biologically produced RNA have so far failed to provide concrete proof regarding the RNA world being the pathway between non-life and life.To date, no possible explanation has been advanced as to how primitive self – replicating RNA molecules could have made transition into modern cellular systems that rely heavily on a variety of proteins to process genetic information. The RNA World hypothesis that hints at first living organism having an RNA-based genome appears to be on shaky ground also considering reproductive strategy of viruses whose RNA-based genome requires DNA of a living host to survive and replicate. Moreover, despite development of sophisticated biotechnology tools in the recent years, scientists still have not been successful in transforming inanimate matter into life in the laboratory. The available scientific knowledge simply fails to provide the viable clues about the process of evolution of life from non-life. Besides, the underlying uniformity of life on the Earth, with all modern organisms sharing the same DNA-based mechanism for genetic transmission, is indicative of the fact that life emerged here only once during the planet's entire history (Burliinski 2006).It is ironicthat the crucial timing hitherto remains unknown to the mankind.

Ever since Oparin and Haldane initiated the modern theory of life's origin from non-life in 1930s, we have learnt much about how life operates, but almost nothing about how it originates. It is a puzzle whose mystery will perhaps remain unknown to the humanity *ad infinitum*.

……………………………………………………………………

Maher, K.A. , Stevenson, D.J. (1998). Nature. **331**(6157), 612-614

Ma, Ying-Juan, et al. (2007). Geophys. Res. Lett. **34** (24), L24S10

Manning, Craig E. et al.(2006). American Journal of Science. **306**, 303-306

Meierhenrich, U.J. (2004). Proc. National Academy of Sciences. **101**, 9182-9186

Martin, W. and Russel, M.J. (2002). Philosophical Transactions of Royal Society: Biological Sciences. **358,** 59-85

Mastrapaa, R.M.E. (2001). Earth and planetary Science Letters.**189** (30), 1-8

Matthews, C.N. (1991). Origin of Life and Evolution of Biosphere. **21**(5), 421-434

Miller, Stanley L. (1953). Science. **117**, 528-529

Miller, Stanley L. and Orgel, Leslie E. (1974).'The Origins of Life on the earth', Prentice Hall, New Jersey

Mileikowsky, C. et al (2000). Icarus. **145**, 391

Minard, R.D. et al. (1998). The Journal of International Society for the Study of Origin of Life. **28** (4-6), 461-473

Mondigan, Michael T. and Marrs, Barry L. (1997). Sci. Am. **276**, 82-87

Mojzsis, S.J. et al. (1996). Nature. **384,** 55-59

Mojzsis, S.J. et al. (2001). Nature. **409**,178,

Moore, M.H. and Donn, B. (1982). Astrophysical Journal. **257**, L47- L50

Narlikar, J.V., Lloyd, D., and Wickramasinghe, N.C. (2003). Astrophysics and Space Science. **285** (2), 555-562

Nemchin, A.A. et al. (2008). Nature. **454**, 92-95

New Organic Molecules in Space, http://www.astronomy.com , dt. 28 March 2008

Nicholson, W.L. et al. (2000). Microbiology Molecular Biology Reviews. **64**, 548-572

Ogasawara, H. et al. (2000). Origins of Life and Evolution of the Biosphere. **30**, 527-537.

O' Neil, J. et al. (2008). Science. **321**, 1821-1831

Oparin, A.I (1957). 'The Origin of Life on Earth'(3rd edition). Macmillan, New York
23

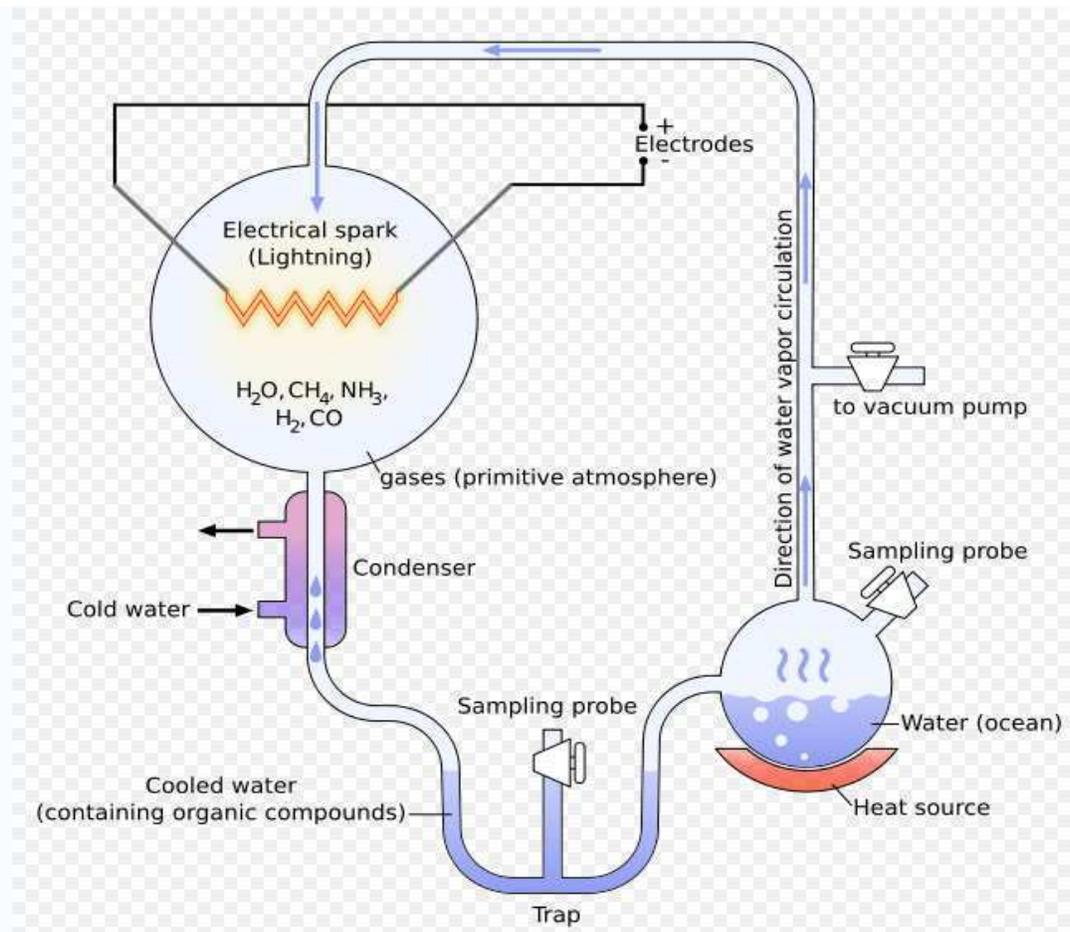

**Fig. 1   Miller – Urey Experiment**

**Adapted from : http://en.wikipedia.org/wiki/Miller-Urey_experiment**

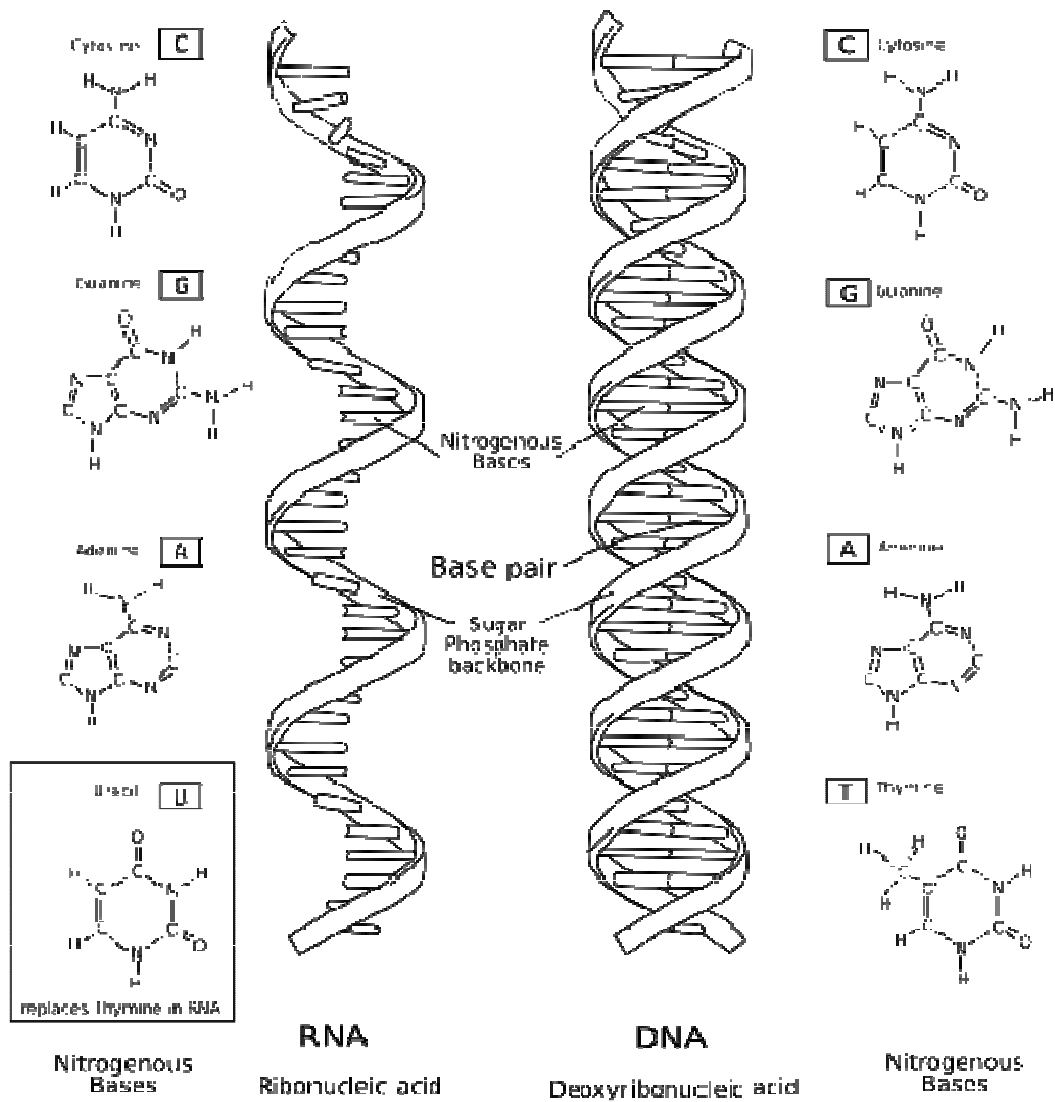

**Fig. 2   RNA with nitrogenous bases to the left and DNA to the right**

**Adapted from :   http://en.wikipedia.org/wiki/RNA_world_hypothesis**

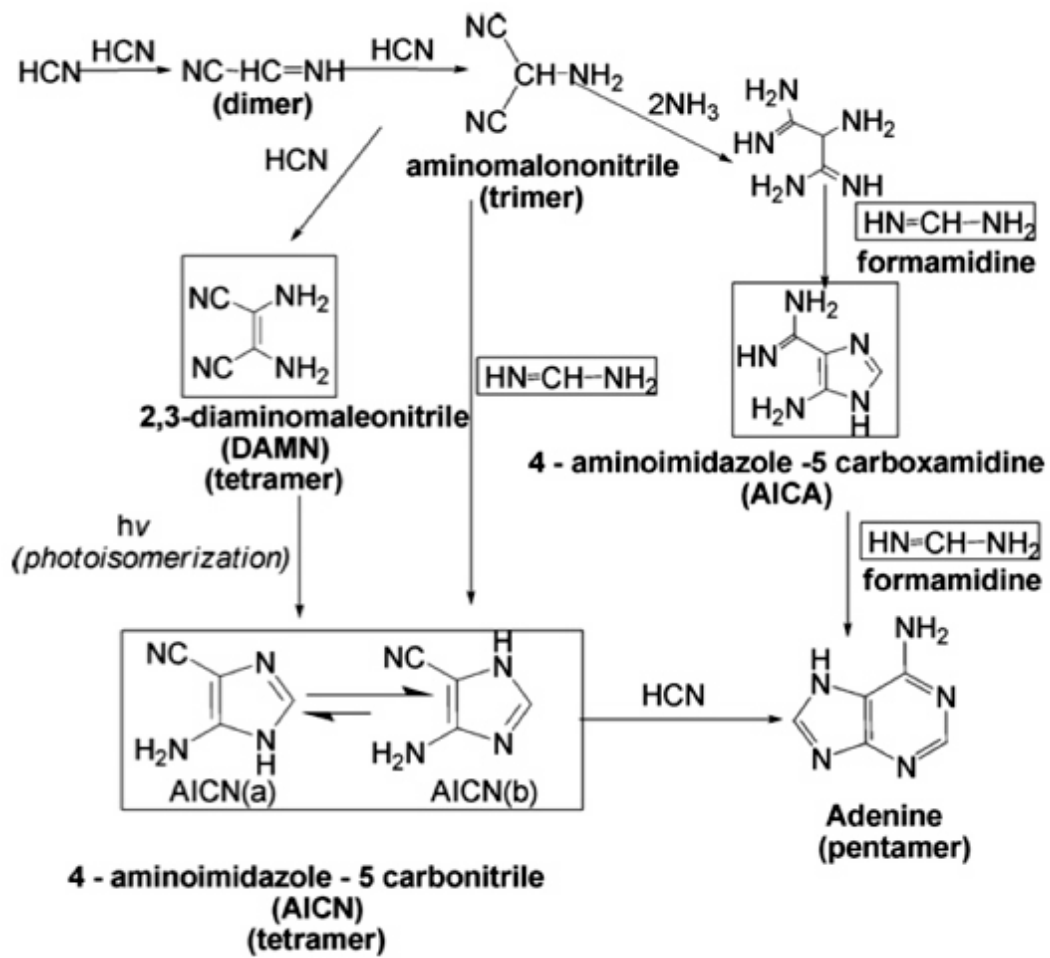

Fig. 3  Proposed steps in the formation of adenine from HCN in aqueous solution

Adapted from : Roy et al. (2007). PNAS, 104(44), 17272-17277

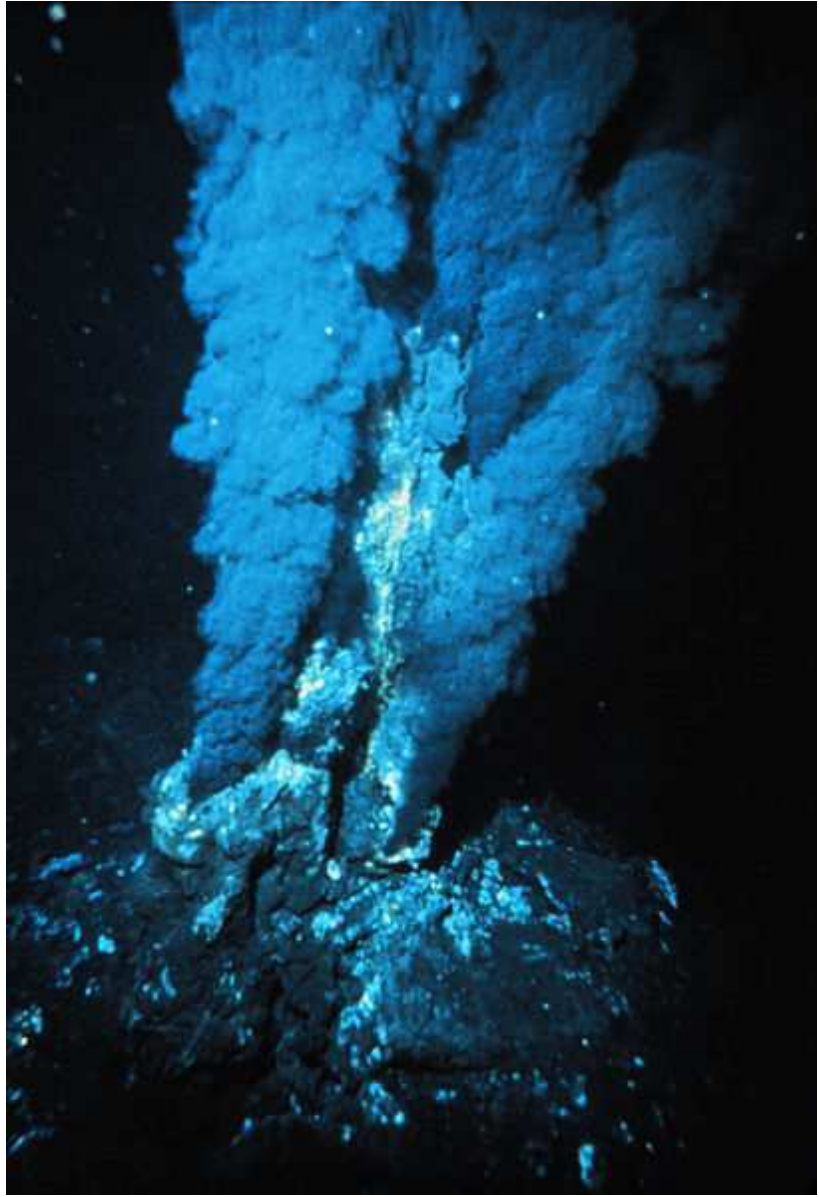

**Fig. 4  Black smoker at mid-ocean ridge hydrothermal vent in the Atlantic ocean**

**Adapted from  : http://www.photolib.noaa.gov/htmls/nur04506.htm**

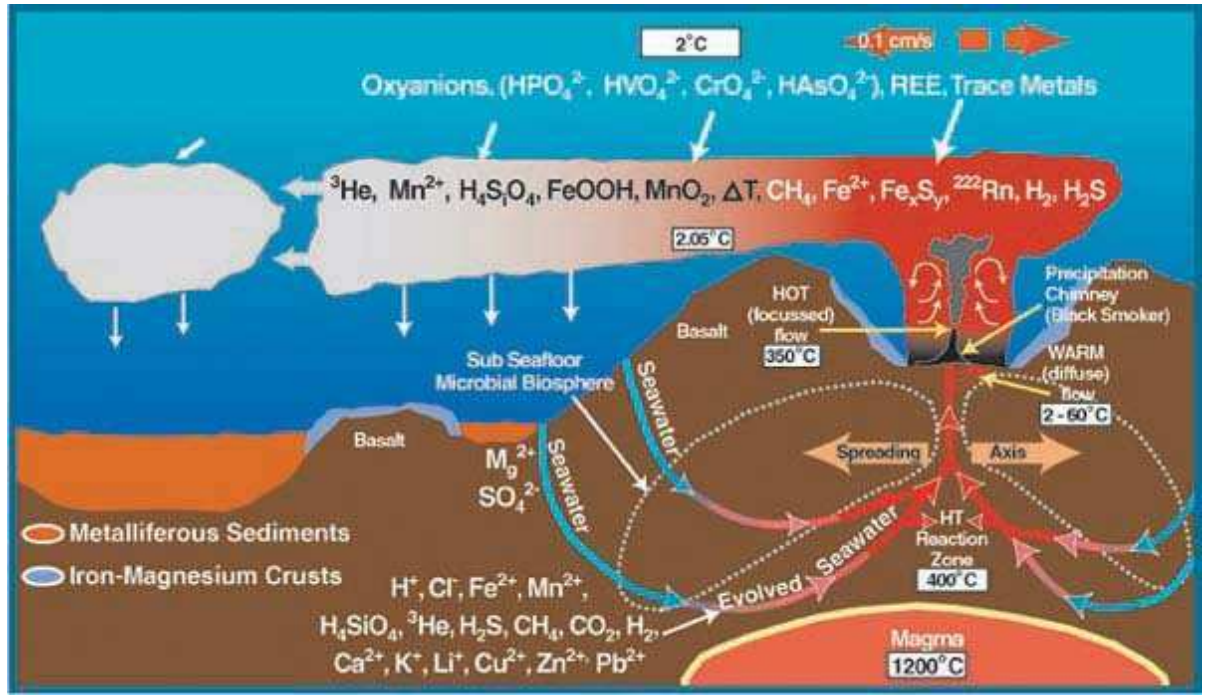

**Fig. 5  Deep sea vent biogeochemical cycle diagram**

Adapted from :

http://en.wikipedia.org/wiki/Image:Deep_sea_vent_chemistry_diagram.jpg

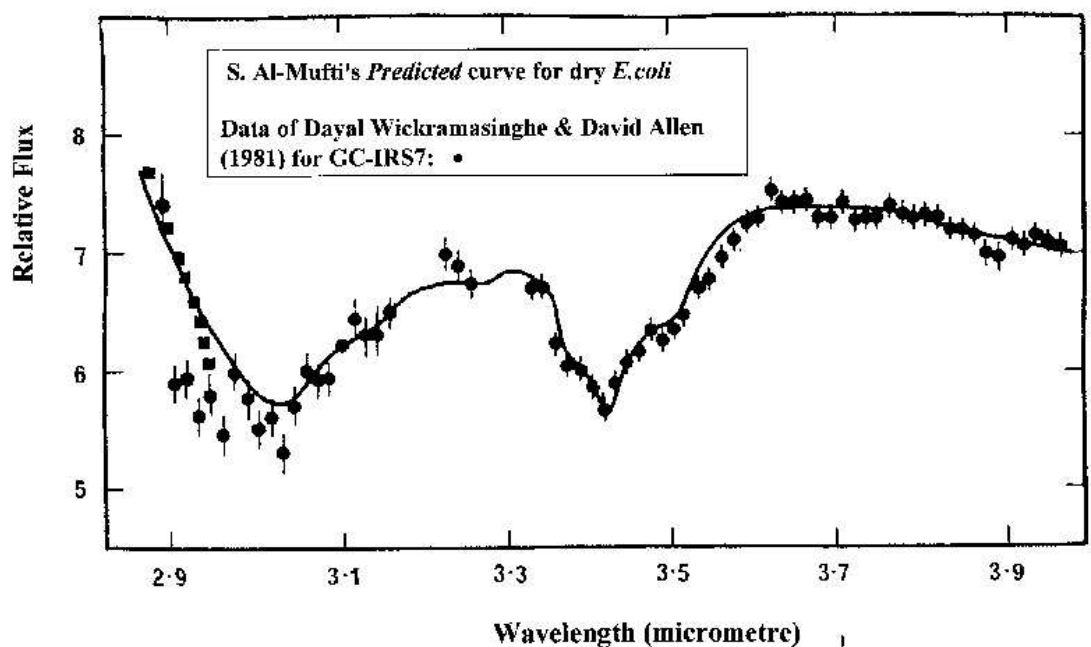

**Fig. 6 The original data for GC-IRS7 and the predicted behaviour of freeze dried bacteria**

**Adapted from : Wickramsinghe, et al. : Proc. SPIE, 4859 , 154 -163, 2003**

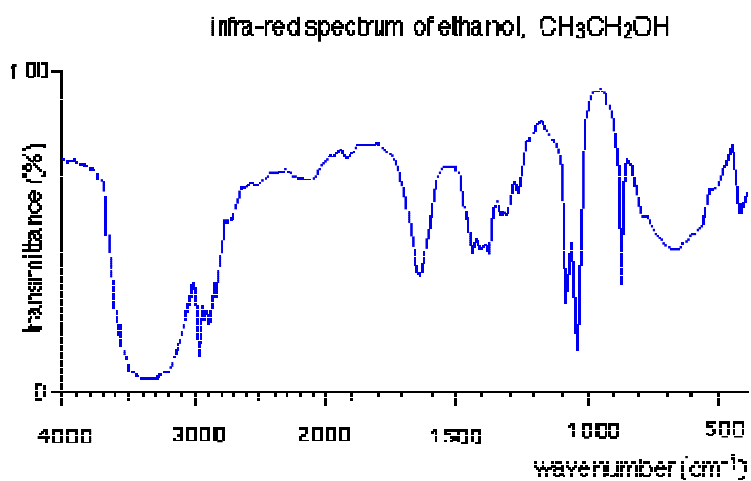

**Fig. 7 Near Infrared spectrum of of ethanol**

**Adapted from : http://www.chemguide.co.uk/analysis/ir/interpret.html#top**

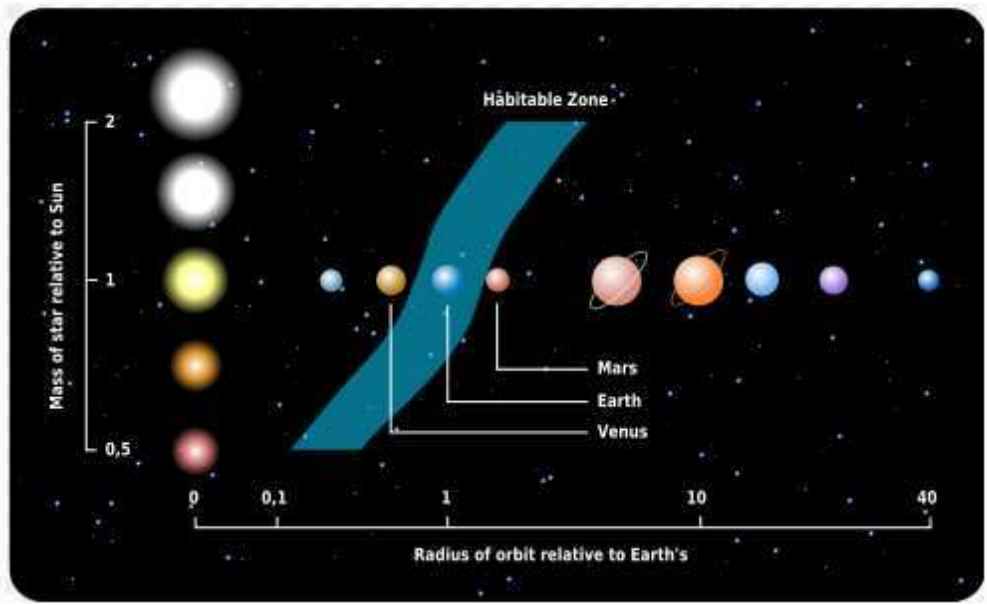

**Fig. 8 Planetary habitability chart showing where life might exist on extrasolar planets based on our own Solar System and life on Earth**

Adapted from : http://en.wikipedia.org/wiki/Image:Habitable_zone-en.svg